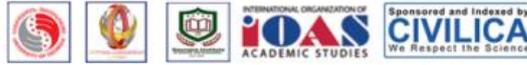

# Optimization of Spectral Efficiency in Cell-Free massive MIMO Systems Using Deep Neural Networks

**Marzieh Arasteh**
Faculty of Electrical Engineering
K. N. Toosi University of Technology
Tehran, Iran
m.arasteh91@email.kntu.ac.ir

**Narges Yarahmadi Gharaei**
Faculty of Computer Engineering
K. N. Toosi University of Technology
Tehran, Iran
yarahmadi@email.kntu.ac.ir

**Mehrdad Ardebilipour**
Faculty of Electrical Engineering
K.N. Toosi University of Technology
Tehran, Iran
mehrdad@eetd.kntu.ac.ir

## Abstract

Cellular communication is a widely used technology in the world where the coverage area is divided into multiple cells. Interference is one of the most important challenges in cellular networks which causes problems by reducing the quality of the service. Cell-Free (CF) massive multiple-input multiple-output (MIMO) is a novel technolgy in which a large number of distributed access points (APs) are concurrently serving a small number of user equipment (UE). CF network can be an alternative technology to cellular networks for reducing interference. A challenging task in a CF network is scalability, where although the number of UEs tends to infinity, the computational complexity must remain finite in each AP or UE. In this paper, we provide two architectures of Dense fully connected neural network (Dense_Net) and 1D convolution neural network (Conv_Net) to be implemented in different cases in terms of the number of antennas in each AP/UE and the method for the combining vector. The Dense_Net outperforms the Conv_Net in all the cases. For instance, in the first case it has a %62.87 improvement in terms of loss. The results show that our proposed method performs better in terms of obtaining optimal values for spectral efficiency (SE).

## Introduction

Massive multiple-input multiple-output (MIMO) has been regarded as a fundamental technology for fifth-generation (5G) and beyond 5G systems [1]. Using multiple antennas at the transmitter and/or receiver can help to improve the energy efficiency (EE) and spectrum efficiency (SE). In conventional cellular massive MIMO systems, macro-diversity was not provided perfectly for reducing the adverse shadowing effects. Moreover, these systems were not capable of supplying a uniform service for all the users in the whole coverage area [2].

Due to the interference problems that exist in most cellular networks currently in use, recently an encouraging technology has been emerged named cell-free (CF) massive MIMO, where a large number of geographically distributed access points  (APs) are jointly serving a small number of user equipment (UE) by joint coherent transmission and reception among APs. The main purpose of designing a CF massive MIMO system was to provide a uniformly high service quality in the entire coverage area. CF networks are divided into the core and the edge. The edge consists of the APs and central processing units (CPUs), which are responsible for the APs cooperation. Connections between APs and CPUs are made via fronthaul links. Sharing the physical-layer signals transmitted in the downlink and forwarding the received uplink signals which need to be decoded are both done by fronthaul links. The edge and the core parts are connected through backhaul links [3].

Since the geographical coverage area of the CF network can be huge, the technology must be scalable. In a scalable network, more APs and/or UEs can be added to the network; however, the capacity of the available ones does not increase. The early CF network papers assumed that all the UEs are served by all the APs. This assumption complicates scalability along with the fact that there is no cell boundary in this type of network.

In the recent past years, deep learning (DL) has become a powerful tool to extract the pivotal features in communication systems and can achieve unrivaled levels of accuracy. A convolutional neural network (CNN) is a DL algorithm that contains three main types of layers, including convolutional layer, pooling layer, and fully connected (FC) layer. CNN is an efficient approach in image processing, but it can also be used in communications [4] such that the channel matrix between the APs and UEs can be assumed as an image, where each element in the matrix denotes a pixel in the image. In image processing, a pixel is the smallest measurable unit. In massive MIMO systems, the computational complexity corresponding to channel estimation, power allocation, or SE optimization increases at least linearly with the number of antennas per user and/or AP. The complexity discussed above includes the operation of multiplying, inversing or transposing large-scale matrices. Not only the DL algorithms can be used to reduce the computational burden, but also they can reduce the elapsed time to achieve the desired output [5].

The structure of the paper is as follows: we refer to the related works in section II and explain the problem in section III. In section IV, preparing the dataset is organized. Dataset preprocessing is explained in section V. Section VI is dedicated to explaining the proposed method. Section VII shows the numerical results. Finally, section VIII presents the conclusion.

## Related Works

In this section, some of the recent developments in CF massive MIMO and deep learning in communication systems are explained.

Luo et al in [6] proposed a two-step offline-online method to predict the channel state information (CSI) from the historical data in 5G communication systems. Its neural network architecture was a combination of 2D CNN, 1D CNN, and a long short term memory (LSTM) which achieves high accuracy for new data in the online step.

In [7], deep learning was used in power allocation problems in  massive MIMO systems.  Due to the  computational complexity in optimizing the received power by each user, DL was a solution to allocate the optimized power for each UE in a cellular network.

In [8], a comparison has been made between a small cell where in each cell the AP serves only one user and a CF massive MIMO in both correlated and uncorrelated shadow fading conditions. Although small cell systems require fewer backhaul resources, they are weak in terms of throughput.

Jin et al in [9] compare the conventional channel estimation methods with capabilities in deep neural networks. At first, it describes the conventional methods like least squares (LS) and minimum mean square error (MMSE) positives and negatives. LS is an estimator in which the achieved normalized mean square error (NMSE) is poor, but its computational complexity is low. However in MMSE, the computational burden is heavy but its achieved NMSE is more acceptable. After studying the traditional methods for channel estimation, an innovative method using neural networks is described.  It provides an algorithm using CNNs for channel estimation in CF millimeter-wave massive MIMO systems.

**System Model**

In this paper, we propose a CF massive MIMO system that consists of P number of geographically distributed APs, each containing N antennas and U number of UEs. Our designed system uses time division duplexing (TDD) technique where, by using synchronized time intervals, the uplink and downlink transmissions are carried over the same frequency. In comparison to the frequency division duplexing (FDD), which uses two different radio carrier frequencies, TDD has much better spectral efficiency (SE) [10]. Moreover, the equipment cost in TDD is more affordable compared to FDD since there is no need to isolate the transmitter and receiver by a diplexer. In a communication system, the time-frequency resources are divided into multiple channel coherence blocks. According to Figure.1, in a TDD protocol the uplink and downlink transmissions use the same coherence bandwidth (BW). The uplink transmission is divided into two phases: the UL pilot phase and the UL data phase. The AP must estimate the channel based on the received pilot signal sent by its serving UE.

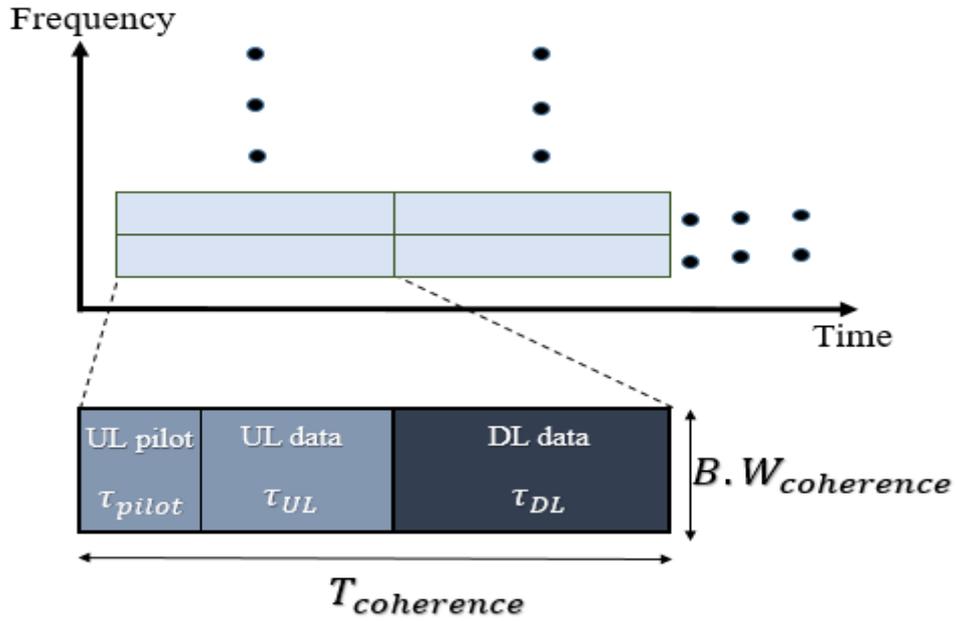

**Figure 1. A TDD protocol in a coherence block**

Based on how the APs are serving the UEs, the networks can be divided into two main categories. The first one is Network Centric Clustering (NCC) where the APs are performing in separated clusters and each UE is only served by the APs in that cluster. This implementation is not efficient for the UEs which are close to the edge of the cluster. The second one is dynamic cooperation clustering (DCC) where the network's operation depends on how the APs select a set of desired UEs to serve. This is the approach used in this paper.

$$\chi_u \subset \{1,2, \ldots, P\} \tag{1}$$

In (1), $\chi_u$ is the subset of APs which serve the $u^{th}$ UE. To be more exact, we introduce a matrix to show the connections between the APs and UEs in a DCC framework.

$$G_{up} = \begin{cases} I_N & p \in \chi_u \\ 0_{N \times N} & p \notin \chi_u \end{cases} \tag{2}$$

In (2), $G_{up}$ is an identity matrix when the $p^{th}$ AP is in the subset of the APs which can transmit and receive signals from the $u^{th}$ UE and a zero square matrix otherwise for $p = 1,2,\ldots,P$ and $u = 1,2,\ldots,U$.

Based on [11], in a scalable CF network, where the number of UEs tends to infinity, the following items must remain executable at each AP:

- Optimization of the power allocation: each AP in the network must perform an optimization algorithm to allocate its transmit power between all the UEs that it serves. Regardless of the optimization type, the computational complexity of optimizing the power allocation increases as the number of UEs→ ∞.

- Signal processing for data transmission and channel estimation: the downlink signal transmitted by the p[th] AP to the UEs which it serves is

$$s_p = \sum_{j=1}^{U} \boldsymbol{G}_{jp} \, \boldsymbol{\beta}_{jp} \, \vartheta_j \tag{3}$$

where $\boldsymbol{G}_{jp}$ determines whether the j[th] AP is communicating with the p[th] UE or not. $\boldsymbol{\beta}_{jp}$ and $\vartheta_j$ denote the transmit precoding vector and unit power data signal. As U→ ∞, the processing operation for the AP requires infinite complexity. On the other hand, the computational complexity for each AP to estimate all the UEs channels which it serves, grows at least linearly with the number of users [12].

- Fronthaul signaling for data sharing: the received signals from the CPU at the AP, must be sent to the serving UEs. Thus, this transmission is proportional to the number of UEs. By increasing the number of UEs to infinity, any memory capacity to store the data of the serving UEs is inadequate.

In [11], to study scalability, we first design the network in such a way that the maximum number of UEs that any AP can serve in the network is unit. Then, based on how the UEs are distributed over the coverage area, one of the neighboring APs is set as the master AP. Master AP must be the most spatially correlated one to the UE. In order to reduce the pilot contamination, the master AP responds to the UE in the way that the minimum amount for the received pilot signal's correlation matrix is obtained. Then, this master AP asks the other APs to see whether they can serve this UE or not. This procedure is feasible only if the neighboring APs do not serve any UE on the same pilot. By increasing the number of the UEs to infinity, the computational complexity and resource requirements remain still finite.

The received signal at the AP must be estimated. This estimation process can be done by applying a receive combining vector to the received uplink signal. Many algorithms for combining vectors have been developed so far. Here, maximum ratio (MR) and local-partial minimum mean square error (LP-MMSE) combining vectors are examined.

$$\boldsymbol{w}_{up}^{MR} = \sqrt{\frac{\rho_{up}}{\mathrm{E}\{|\widehat{H}_{up}|^2\}}} \, \widehat{H}_{up} \tag{4}$$

$$\boldsymbol{v}_{up}^{LP-MMSE} = p_u (\sum_{j \in G_p} p_j (\widehat{H}_{jp}\widehat{H}_{jp}^H + \boldsymbol{C}_{jp}) + \sigma^2 \, \boldsymbol{I}_N)^{-1}\widehat{H}_{up} \tag{5}$$

In (4), MR combining vector is denoted which in known as a low computational complexity method. $\rho_{up}$ is a positive valued number to demonstrate the amount of power that the p[th] AP assigns to the u[th] UE and $\widehat{H}_{jp}$ is the estimated channel matrix between the j[th] UE and p[th] AP.

In (5) from [11], LP-MMSE is a combining vector for the cases where the network is DCC. $p_j$ is the j[th] UE's transmit power and $\boldsymbol{C}_{jp}$ is the collective estimation error. The combining operation in LP_MMSE only requires the estimates of the channel. Moreover, the number of complex multiplications is proportional to the number of antennas at each AP, and the communication between the u[th] UE and p[th] AP.

$$\mathcal{N} \propto \{N, |\boldsymbol{G}_{up}|, |\chi_u|\} \tag{6}$$

According to the number of complex multiplications in (6), LP-MMSE is a scalable issue for combining vector in a CF massive MIMO system because the number of complex multiplications in the combining vector doesn't relate to the number of UEs. Thus, it is possible to let U→ ∞. In order to calculate the SE in [11], multiple parameters are needed to be given as input which leads to an increase in computational complexity and run time. These parameters include the channel matrix and the estimated channel matrix between the UEs and APs serving them, spatial correlation matrix, length of the coherence block and pilot sequence, downlink transmit power matrix that each AP allocates to its serving UEs when using distributed precoding. As is clear, the computation of SE in a scalable network depends on numerous parameters which increases the bulk of software computations.

**Preparing Dataset**

In this section, simulation results are explained in detail to perform a quantitative comparison. Our network coverage simulation environment is a square with an area of 4 km². The communication bandwidth is 20 MHz. The threshold value is -40 dB when a UE will be served by an AP that is not its Master AP. For the channel's matrix, we first consider an initial $H_{P \times O \times U}$ matrix which is generated randomly and contains both real and imaginary parts. Each of these parts has three dimensions. The first dimension denotes the number of the APs, the second dimension represents the number of the realizations in which the channels are being realized and the third one stands for the number of the UEs. The channel is estimated through the network and a new estimated channel matrix is obtained using MMSE estimator which is indicated by $\widehat{H}_{P \times O \times U}$. $P, O$ and $U$ stand for the number of APs, realizations and UEs. In this paper, three cases are compared to each other which consist:

- Case1: the network is not DCC. All the APs are serving the whole number of UEs. The matrix $G_{up}$ is 1 for all the elements. The amounts of APs, antennas per each AP, and the number of UEs in this case are 256, 1, and 80, respectively. The combining vector method is MR and the network is not scalable.
- Case2: the network is considered to be scalable and the combining vector is LP-MMSE. Moreover, the network is DCC, where each UE is served only by some preferred APs in its neighboring. The number of APs, antennas per each AP, and the UEs are the same as case1.
- Case3: the last case is similar to case 2 in scalability and combining vector operation, but the amounts of APs, N, and UEs are equal to 32,8,80.

As is evident, the total number of AP antennas which is obtained by $L \times N$ equation, is constant equal to 256 in all the above cases. The simulations occur in 12 Monte Carlo setups, which is a high-performance computational algorithm. It uses random sampling repeatedly in order to solve the various problems.

**Dataset Preprocessing**

By performing various experiments, in our method only the two main parameters which are recognized as more important than the others are examined according to Figure. 2: The $G_{P \times U}$ and $\widehat{H}_{P \times O \times U}$. The dataset of these two parameters in three cases described above, are obtained and given as the input to two different neural networks: a dense network and a CNN network.

The network's input contains three main parts: the real part of the $\widehat{H}_{P \times O \times U}$ matrix, the imaginary part of the $\widehat{H}_{P \times O \times U}$ matrix and the $G_{P \times U}$ matrix. In each realization, a $P \times U$ matrix in both the real and imaginary parts of the $\widehat{H}_{P \times O \times U}$ which correspond to the $G_{P \times U}$ matrix are given to the neural networks. The output is a vector of 12 elements which denotes the number of the Monte-Carlo setups. In the training progress, depending on the type of the combining vector in each case described previously, different outputs SE matrices are dedicated.

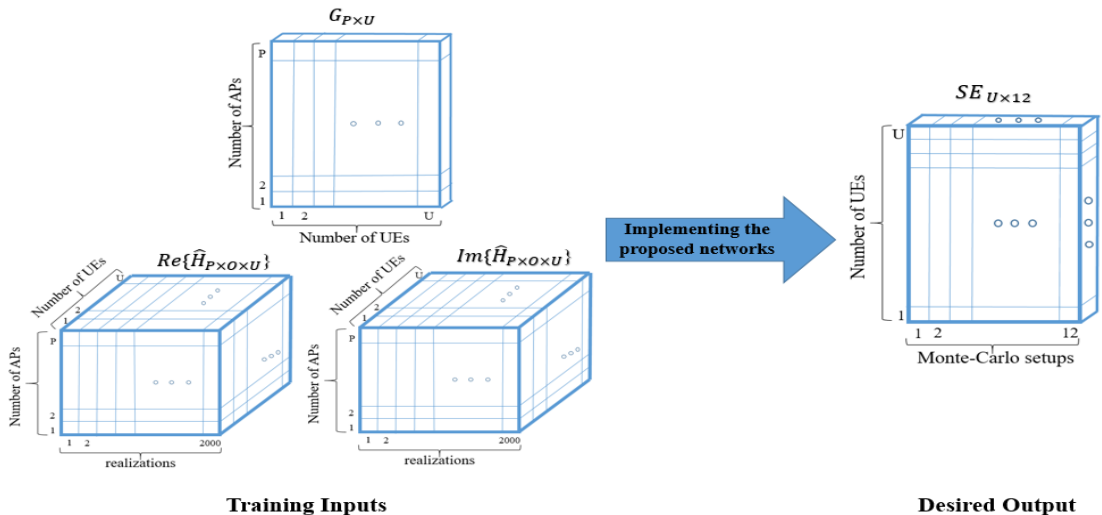

**Figure 2.** Visualizing the training process using the inputs and the desired output

**Proposed Method**

Nowadays, the usage of neural networks and deep learning is embedded in many scientific and practical fields. Among the application of machine learning, we can mention intelligent reflecting surface (IRS) [13], non-orthogonal multiple access (NOMA) [14] related to communication systems and predicting the production rate related to oil field [15] and business [16].

A fully connected neural network is a type of neural network in which all neurons in a layer are connected to all neurons in the next layer. A dense layer is one of the most common neural network layers that is connected deeply which each neuron of this layer receives all outputs of all neurons in the previous layer. A convolutional neural network (CNN, or ConvNet) is a class of deep neural networks mostly used to analyze images and videos such as [17]. Each convolution layer convolves the input by a specific filter and passes its result to the next layer [18]. Time-Series data or signals are used as one dimensional CNN [19]. In Conv1D, the kernel slides along one dimension.

To find the best value of the SE vector, we have proposed two methods. The first method is a fully connected dense neural network called as Dense_Net and the second one is a convolutional neural network called as Conv_Net. In the following, we will describe these two architectures.

**Numerical Results**

As shown in Figure. 3, the architecture of the proposed Dense_Net is visualized. Figure. 4 shows the architecture of Conv_Net. These two architectures are applied to all of the three cases explained in preparing dataset under a fair condition. We have split the dataset into two parts, 80 percent for the training process and 20 percent for the testing process. Also, during the training process, the data is split by a rate of about 0.03125 for validating the model. The training is done with 150 epochs. Our loss function is the performance of the model monitored by the Root Mean squared Error (RMSE) value. In this regard, the model was trained to achieve the minimum RMSE. We minimized the error by Adam's [20] optimization algorithm. This model is trained four times sequentially to find the best global minimum with learning rates of 0.001, 0.0001, 0.0001, and 0.00001, and beta_1 equal to 0.9, beta_2 equal to 0.999. The mentioned alpha_1 and alpha_2 gave us the best output among the various tests.

```
Size input vector (409600, 240)
Size output vector (409600, 12)
Model: "sequential"

_________________________________________________________________
 Layer (type)                Output Shape              Param #
=================================================================
 layer1 (Dense)              (None, 128)               30848

 layer2 (Dense)              (None, 64)                8256

 layer3 (Dense)              (None, 32)                2080

 layer4 (Dense)              (None, 32)                1056

 layer6 (Dense)              (None, 24)                792

 layer7 (Dense)              (None, 12)                300

=================================================================
Total params: 43,332
Trainable params: 43,332
Non-trainable params: 0
```

**Figure 3. Architecture of Dense_Net**

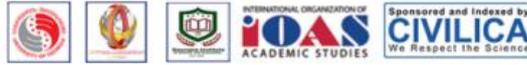

```
Size input vector (409600, 240, 1)
Size output vector (409600, 12)
Model: "sequential_1"

_________________________________________________________________
 Layer (type)                Output Shape              Param #
=================================================================
 conv1d_2 (Conv1D)           (None, 238, 20)           80

 conv1d_3 (Conv1D)           (None, 236, 20)           1220

 dropout_1 (Dropout)         (None, 236, 20)           0

 max_pooling1d_1 (MaxPooling  (None, 118, 20)          0
 1D)

 flatten_1 (Flatten)         (None, 2360)              0

 dense_2 (Dense)             (None, 20)                47220

 dense_3 (Dense)             (None, 12)                252

=================================================================
Total params: 48,772
Trainable params: 48,772
Non-trainable params: 0
```

**Figure 4. Architecture of Conv_Net**

We have implemented our design model with Keras on GPU of google Colab with 12 Gigabyte RAM. The loss and validation loss and run time of each implementation are visible in TABLE I.

According to the results shown in TABLE I, Dense_Net performed better than Conv_Net in all three cases. Also, the training time of Dense_Net is less than Conv_Ne. As it is obvious, based on the fact that the amount of loss is very small, it can be concluded that Dense_Net produces the desired output very accurate, which is a very significant point. Combining vectors in conventional methods including MR and LP-MMSE need to receive multiple parameters as input to obtain the SE in the uplink, whereas in Dense_Net and Conv_Net it is only important to know how the APs and UEs are connected to each other and the channel statistics as input.

**TABLE I. Results of all implementations**

| Model Architecture | Case | Loss | Validation loss | Run time (s) |
|---|---|---|---|---|
| Dense_Net | 1 | 0.0738 | 0.0764 | 2200.198 |
| | 2 | 1.0120e-05 | 1.0805e-05 | 2256.598 |
| | 3 | 1.6940e-05 | 1.7983e-05 | 495.414 |
| Conv_Net | 1 | 0.1202 | 0.0921 | 6571.12 |
| | 2 | 0.078 | 0.0567 | 6278.454 |
| | 3 | 0.0124 | 0.0131 | 1997.703 |

In Figure. 5, the network is not scalable with MR combining vector. The number of APs, antennas per each AP and UEs is 256,1 and 80. In this case, Dense_Net and Conv_Net outperform the conventional MR combining vector method because the curve of both networks have moved far to the right.

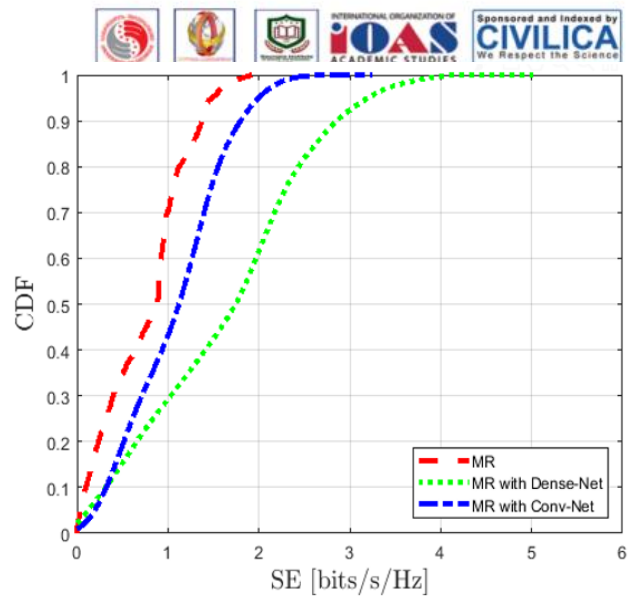

**Figure 5.  CDF of the uplink SE with P=256 APs, N=1 antenna and U=80 with non-scalable combining vector**

Figure. 6 shows that the Conv_Net and Dense_Net have a significant improvement over the conventional LP-MMSE method for the achieved uplink SE. In this case, the network is assumed to have 256, 1 and 80 number of APs, antennas per each AP and UEs.

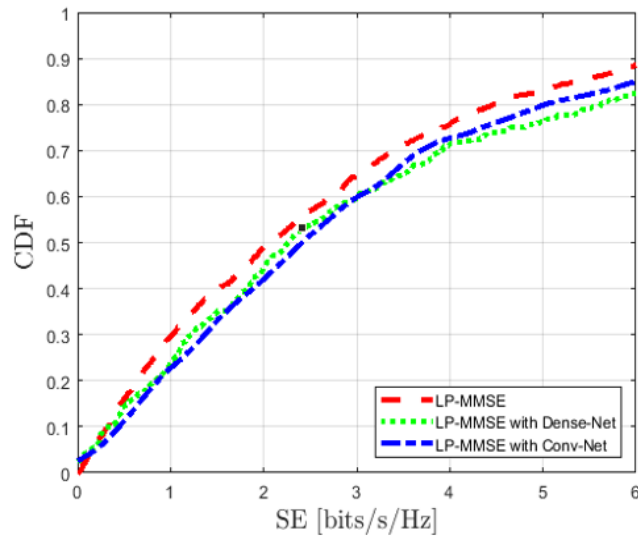

**Figure 6.  CDF of the uplink SE with P=256 APs, N=1 antenna and U=80 with scalable combining vector**

In Figure. 7, the network is scalable and the number of APs, antennas per each AP and UEs is equal to 32, 8 and 80. In comparison to the previous cases, Dense_Net and Conv_Net have the least impact on the SE performance. However, negligible improvements have been made in obtaining the optimal SE values.

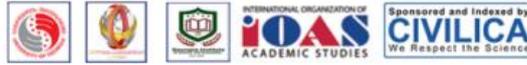

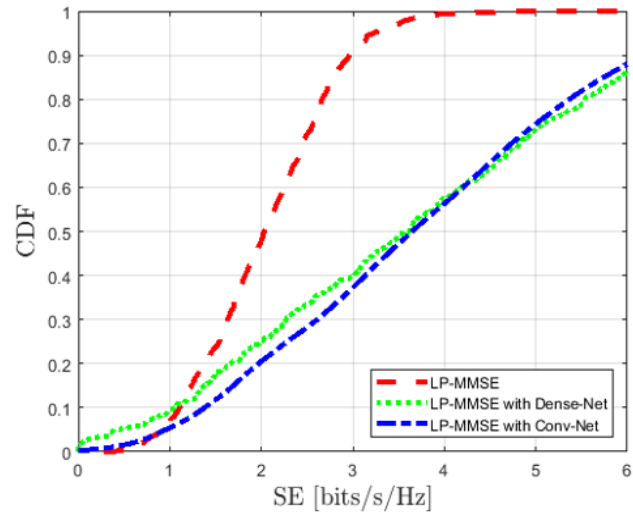

**Figure 7. CDF of the uplink SE with P=32 APs, N=8 antennas and U=80 with scalable combining vector**

**Conclusion**

In this paper, we proposed two DL networks to achieve optimal SE in a scalable CF massive MIMO system, where the computational complexity was significantly reduced compared to traditional methods. The inputs and output were previously generated as the dataset. The number of inputs to obtain the optimal SE were reduced to two major ones. The networks were examined in three cases using different scalable and non-scalable combining vectors and different number of antennas per each AP.

Results show that the achieved SE in our model has considerable improvement in all of the three cases.

**Acknowledgment**

This publication was supported by grant No. RD-51-9911-0024 from the R&D Center of Mobile Telecommunication Company of Iran (MCI) for advancing information and communications technologies. We thank Dr. Ali Rakhshan for his comments that improved the manuscript.

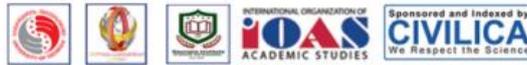